\documentclass{elsart}
\usepackage{graphics,natbib,graphicx,psfig,amssymb,ccaption} 
\usepackage{lscape} 
\journal{New Astronomy}
\input psfig.sty
\begin{document}
\begin{frontmatter}
\title{The age of cataclysmic variables: \\ a kinematical study}
\author[istanbul,tug]{T. Ak\corauthref{cor}},
\corauth[cor]{corresponding author.}
\ead{tanselak@istanbul.edu.tr or tanselak@tug.tubitak.gov.tr}
\author[istanbul]{S. Bilir},
\author[istanbul]{S. Ak},
\author[istanbul]{K.B. Co\c skuno\u glu},
\author[tug,akdeniz]{Z. Eker}
\address[istanbul]{Istanbul University, Faculty of Sciences, Department 
of Astronomy and Space Sciences, 34119 University, Istanbul, Turkey}
\address[tug]{T\"UB\.ITAK National Observatory, Akdeniz University 
Campus, 07058 Antalya, Turkey} 
\address[akdeniz]{Akdeniz University, Faculty of Arts and Sciences, Department of 
Physics, 07058, Campus, Antalya, Turkey}

\begin{abstract}

Using available astrometric and radial velocity data, the space velocities 
of cataclysmic variables (CVs) with respect to Sun were computed and kinematical 
properties of various sub-groups of CVs were investigated. Although observational errors of systemic velocities ($\gamma$) are high, propagated 
errors are usually less than computed dispersions. According to the analysis of propagated 
uncertainties on the computed space velocities, available sample is refined by 
removing the systems with the largest propagated uncertainties so that the 
reliability of the space velocity dispersions was improved. 
Having a dispersion of $51\pm7$ km~s$^{-1}$ for the space velocities, 
CVs in the current refined sample (159 systems) are found to have $5\pm1$ Gyr mean 
kinematical age. After removing magnetic systems from the sample, it is found that 
non-magnetic CVs (134 systems) have a mean kinematical age of $4\pm1$ Gyr.
According to $5\pm1$ and $4\pm1$ Gyr kinematical ages implied by 
$52\pm8$ and $45\pm7$ km~s$^{-1}$ dispersions for non-magnetic 
systems below and above the period gap, CVs below the period gap are older than 
systems above the gap, which is a result in agreement with the standard evolution theory 
of CVs. Age difference between the systems below and above the gap is smaller than 
that expected from the standard theory, indicating a similarity of the angular 
momentum loss time scales in systems with low-mass and high-mass secondary 
stars. Assuming an isotropic distribution, $\gamma$ velocity dispersions of non-magnetic CVs 
below and above the period gap are calculated $\sigma_\gamma=30\pm5$ km s$^{-1}$ and 
$\sigma_\gamma=26\pm4$ km s$^{-1}$. Small difference of $\gamma$ velocity dispersions 
between the systems below and above the gap may imply that magnetic braking does not 
operate in the detached phase, during which the system evolves from the post-common envelope 
orbit into contact.

\end{abstract}

\end{frontmatter}

\section{Introduction}

Cataclysmic variables (hereafter CVs) are short-period interacting binary 
stars consisting of a white dwarf primary and low-mass secondary which overflows 
its Roche lobe and transfers matter to the white dwarf usually via a gas stream and an 
accretion disc. A bright spot is formed in the location where the matter stream 
impacts on the accretion disc. However disc formation is prevented in systems with 
strongly magnetic primaries. Nevertheless mass transfer continues through channels 
as accretion flows. For comprehensive reviews of the properties of CVs, 
see \citet{War95} and \citet{Hel01}.

According to standard CV formation and evolution scenario, CVs are formed from much 
wider -detached- main-sequence binaries with orbital periods typically 10-1000 d of 
the pairs with large mass ratios; secondaries being less than a solar mass, but the 
primaries are more massive ($M_{1}\simeq1-9M_{\odot}$). The primary star evolves in 
a nuclear time scale of roughly $t_{nuc}\simeq10^{10}(M_{1}/M_{\odot})^{-3}$ yr \citep{KS96,Hel01}. 
When the primary fills its Roche lobe, dynamically unstable mass transfer leads to 
a common envelope (CE) phase during which dynamical friction extracts orbital angular 
momentum to eject the envelope of the giant. The time spent in common envelope phase 
is very short compared to other phases of the evolution, of the order of 
$10^{3}-10^{4}$ yr. After the CE phase, system finds itself in a new detached phase 
called post common envelope phase or pre-CV state; the secondary component is still 
a main-sequence star, while the primary component has been changed to a white dwarf. 
This detached system approaches a semi-detached state by either the nuclear expansion 
of the secondary, or by shrinking orbit due to the loss of orbital angular momentum 
by gravitational radiation and magnetic braking. During the post-CE phase, the orbit 
shrinks from the post-CE period of about 0.1-10~d to the orbital period 
$P\simeq9(M_{2}/M_{\odot})$~h at the onset of mass transfer as a CV. The post-CE phase 
can be very short if the system emerges from the CE phase at almost semi-detached 
configuration \citep{KS96}.

The standard scenario proposed for the evolution of CVs after the onset of the mass 
transfer is sensitive to orbital angular momentum loss mechanisms too as in the post CE phase 
of the evolution. It has been proposed two main mechanisms: magnetic braking and gravitational 
radiation. For CVs with orbital periods ($P$) above the period gap, the secondary star is 
partially convective and, thus, it has a magnetic field. It has been proposed that the 
magnetic field in the secondary star causes orbital angular momentum loss by magnetic braking 
\citep{VZ81,Retal82,Retal83,PS83,SR83,K88} until the orbital period decreases to about 3~h 
at which the secondary star becomes fully convective and magnetic braking ceases. Consequently, 
the secondary star relaxes to its thermal equilibrium and shrinks inside 
its Roche lobe. At this point, the mass transfer shuts off completely and the gravitational 
radiation becomes dominant mechanism for the angular momentum loss \citep{Pac67}. Since the 
mass transfer ceased, the system cannot be observed as a cataclysmic variable and it evolves 
towards shorter orbital periods through emission of gravitational radiation. At the orbital 
period of about 2~h, the secondary star starts to fill its Roche lobe again and re-starts the 
mass transfer at a much lower rate than in the mode of long-period CVs. When the orbital 
period decreases to the observed minimum of about 80 min, the secondary star becomes 
a degenerate (brown-dwarf like) object and, thus, further mass transfer causes the orbital 
period to increase \citep{Pat98}, creating systems called period bouncers. For a comprehensive 
review of the formation and evolution of CVs, see \citet{Ritter08}.

Though the above model successfully explains the period gap, i.e. drop in the number of known 
CVs in between orbital periods in the range about $2<P$(h) $<3$, some predictions of it are 
contradictory with observations. Since CVs should spend most of their lifetime near the period 
minimum, a significant accumulation of CVs near the period minimum is expected. Although such 
an accumulation has not been observed for many years \citep{KB99}, there is now a claimed 
discovery of a spike at the period minimum \citep{Southworthetal08,Gaensickeetal09}. According 
to population synthesis \citep{dK92,dKR93,Kol93,Pol96,KB99,Howetal97,KS02}, the vast majority 
of CV population should have orbital periods shorter than about 2~h . However, such an accumulation 
has not been seen in the distribution of orbital periods of CVs. Recently, \citet{Litt08} found 
seven systems with brown dwarf donors and proposed that the missing population of post-period 
minimum CVs has finally been identified, although their masses and radii are inconsistent with model 
predictions. In addition, the predicted space density of CVs is about a few $10^{-5}$ to a few 
$10^{-4}$ pc$^{-3}$ \citep{dK92,Pol96}, whereas the space density derived from the observations is 
in agreement only with the lower limit of these predictions \citep{Pretetal07a,Schetal02,Aketal08}. 
Although a number of alternative suggestions \citep{LP94,KK95,Clemetal98,Koletal98} and some 
alternative angular momentum loss mechanisms \citep{Andetal03,TS01,Scheetal02,Willetal2005,Willetal2007} 
has been introduced as possible solutions to these problems, none of them could solve the 
problem with a complete success. 

The observational data sets, which can be used to test the above predictions of the standard 
evolution model of CVs, are strongly biased by the selection effects. Brightness dependent 
selection effects are the strongest since the known CV sample is not approximately 
magnitude-limited, even if a limiting magnitude as bright as $V=13$ is adopted \citep{Pretetal07b}. 
Nevertheless, kinematical age of a cataclysmic variable, which is a time span since formation of 
component stars, can be used to test the predictions of the model. The age of a CV 
does not affect its mass transfer rate at a given orbital period. Hence the age distribution 
of CVs is not biased by brightness-selection \citep{Kol01}. Nevertheless, shorter the orbital 
period, it is harder to measure radial velocity variation for determining orbital parameters. 

\cite{KS96}, who determined the age structure of a modeled population for galactic CVs by 
applying standard models for the formation and evolution of CVs, predicted that systems above 
the period gap ($P\gtrsim$ 3~h) must have an average age of 1~Gyr, with most of them being 
younger than 1.5~Gyr, while systems below the gap ($P\lesssim$ 2~h) should display a wide range 
of all ages above about 1~Gyr, with a mean of 3-4~Gyr \cite[see also][]{RB86}. Using the 
empirical relation between age $t$ and space velocity dispersion $\sigma(t)$ of field stars 
in the solar neighbourhood \citep{Wie77,Wieetal92}, $\sigma(t)= k_{1}+k_{2}t^{1/2}$ ($k_{1}$ 
and $k_{2}$ are constants), an intrinsic dispersion of the $\gamma$ velocities 
$\sigma_{\gamma} \simeq 15$ km~s$^{-1}$ for the systems above the period gap, and 
$\sigma_{\gamma} \simeq 30$ km~s$^{-1}$ below the gap were predicted by \cite{KS96}. \cite{Kol01} 
states that the difference between ages of systems above and below the period gap is mainly due 
to the time spent evolving from the post-common envelope (post-CE) orbit into contact. 
However, \cite{vPaetal96} (hereafter referred to as vPAS96), who collected the observed $\gamma$ 
velocities for a sample of CVs from published radial velocity studies and analysed this CV sample 
statistically, could not detect such a difference between the velocity dispersions for the systems 
above and below the period gap. Using precise $\gamma$ velocity measurements of only four dwarf 
novae above the period gap, \cite{Noretal02} suggest a velocity dispersion of $\sim$8~km~s$^{-1}$. 
It should be noted that \cite{Kol01} also expresses that if magnetic braking does 
not operate in the detached phase, the $\gamma$ velocity dispersions of CVs are 
$\sigma_{\gamma} \simeq 27$ km~s$^{-1}$ above the period gap versus 
$\sigma_{\gamma} \simeq 32$ km~s$^{-1}$ below the gap.

Aim of this paper is to derive the observed $\gamma$ velocity dispersions to test predictions. 
Kinematical age profiles for CVs according to different orbital period regimes are investigated 
in order to understand orbital period evolution of CV orbits. In order to do this, we have collected 
the measured $\gamma$ velocities of CVs from the published radial velocity studies since the work 
of vPAS96 and combined our $\gamma$ velocity collection with their sample.  

\section{The data}
Distances (or parallaxes), proper motions and $\gamma$ velocities were collected for 194 CVs which 
are listed in Table 1. The columns of the table are organized as name, equatorial ($\alpha, \delta$) 
and galactic ($l$, $b$) coordinates, colour excess $E(B-V)$, type of the CV, and its orbital period. 
The types, equatorial coordinates and orbital periods of CVs were mostly taken 
from \citet[][Edition 7.7]{RK03} and \cite{Downetal01}. 

%Table 1
\begin{table*}
\scriptsize{
\begin{center}
\caption{The data sample. CV denotes CVs with unknown types, DN dwarf novae, NL nova-like stars and N 
novae. The last column is for references. If there is only one number, it refers to colour excess. 
If there are two numbers, the first one is for colour excess and the other for orbital period. 
First five lines of Table 1 are given here. The table can be obtained electronically.}
\begin{tabular}{rlcccccccc}
\hline
 ID & Star & $\alpha$(J2000) & $\delta$(J2000) & $l$ & $b$  & $E(B-V)$   &  Type & $P$ &  Refs \\
    &      & hh:mm:ss & dd:mm:ss & ($^\circ$) & ($^\circ$) &  &  & (d) &   \\
\hline
  1 & AR And & 01:45:03.28 & +37:56:32.7 & 134 &-23 & 0.020 & DN  & 0.1630 & 1 \\
  2 & DX And & 23:29:46.68 & +43:45:04.1 & 108 &-16 & 0.200 & DN  & 0.4405 & 2 \\
  3 & LX And & 02:19:44.10 & +40:27:22.9 & 141 &-19 & 0.032 & DN  & 0.1510 & 3;20 \\
  4 & PX And & 00:30:05.81 & +26:17:26.5 & 117 &-36 & 0.045 & NL  & 0.1464 & 3 \\
  5 & RX And & 01:04:35.54 & +41:17:57.8 & 126 &-21 & 0.020 & DN  & 0.2099 & 1 \\
  . & ...    & ...         & ...         & ... &... &  ...  & ... & ...    & ... \\
  . & ...    & ...         & ...         & ... &... &  ...  & ... & ...    & ... \\
  . & ...    & ...         & ...         & ... &... &  ...  & ... & ....   & ... \\
\hline
\end{tabular}
\\
\end{center}
1) \cite{BruchandEngel1994}, 2) \cite{Drewetal91}, 3) \cite{Aketal08}, 4) \cite{Schmidtetal01}, 
5) \cite{Harretal04}, 6) \cite{Greineretal01}, 7) \cite{Thor04}, 8) \cite{Greineretal99}, 
9) \cite{Baptistaetal00}, 10) \cite{Ladous91}, 11) \cite{Thor86}, 12) \cite{Petersetal05}, 
13 \cite{Wagneretal98}, 14) \cite{Tovmassianetal00}, 15) \cite{Maucheetal94}, 
16) \cite{Froningetal03}, 17) \cite{Weightetal94}, 18) \cite{Thor02a}, 19) \cite{Southworthetal06}, 
20) \cite{Sheetsetal07}, 21) \cite{Rodetal07}, 22) \cite{Withametal2007}, 23) \cite{Tovmassianetal07}     
\\
}
\end{table*}

%Table 2
\begin{table*}
\setlength{\tabcolsep}{4pt}
\tiny{
\begin{center}
\caption{Star, parallaxes ($\pi$), proper motions ($\mu_{\alpha}cos\delta$, $\mu_{\delta}$), systemic 
velocities ($\gamma$) and space velocity components ($U$, $V$, $W$) for the systems listed in Table 1. 
In each row, the first and second references are for the parallax and the proper motion, respectively. 
The third and further references are for $\gamma$ velocity. First five lines of Table 2 are given here. 
The table can be obtained electronically.}
\begin{tabular}{clcccccccl}
\hline
ID  &  Star & $\pi$ & $\mu_{\alpha}cos\delta$ &   $\mu_{\delta}$ &   $\gamma$   &   $U$ &   $V$ &  $W$   &  Refs \\
 &  & (mas) & (mas) & (mas) & (kms$^{-1}$) & (kms$^{-1}$) & (kms$^{-1}$) & (kms$^{-1}$) &   \\
\hline
  1 &     AR And 	  & 3.17$\pm$0.53 & 8.0$\pm$1.0   &-8.0$\pm$2.0   &-5.0$\pm$5.0    &-10.22$\pm$3.65   &-15.43$\pm$4.20  &-6.20$\pm$3.62    &     1;8;16 \\
  2 &     DX And 	  & 1.11$\pm$0.16 &-13.8$\pm$5.4  &-10.8$\pm$5.4  &-1.9$\pm$2.3    & 47.83$\pm$24.42  & 13.72$\pm$9.93  &-23.04$\pm$22.45  &     1;8;17 \\
  3 &     LX And 	  & 3.07$\pm$0.32 &-4.1$\pm$5.9   &-8.8$\pm$5.9   &-48.0$\pm$3.0   & 34.23$\pm$6.63   &-30.49$\pm$7.52  & 1.80$\pm$8.79    &     1;8;18 \\
  4 &     PX And 	  & 1.35$\pm$0.24 &-11.0$\pm$2.7  &-11.2$\pm$2.6  &-27.4$\pm$21.4  & 42.88$\pm$14.42  &-18.3$\pm$16.68  &-12.17$\pm$15.49  & 1;8;19, 17 \\
  5 &     RX And 	  & 5.56$\pm$0.81 & 2.1$\pm$0.7   &-22.5$\pm$1.1  & 0.0$\pm$8.0    &-1.94$\pm$4.37    &-7.10$\pm$6.16   &-17.74$\pm$4.01   &     1;8;17 \\
  ... &     ... 	  & ... & ...   & ...  & ...   &  ...   & ...   & ...   &  ... \\
  ... &     ... 	  & ... & ...   & ...  & ...   &  ...   & ...   & ...   &  ... \\
  ... &     ... 	  & ... & ...   & ...  & ...   &  ...   & ...   & ...   &  ... \\
\hline
\end{tabular}  
\end{center}
1) \cite{Aketal07a}, 2) \cite{Due99}, 3) \cite{Thor03}, 4) \cite{Harretal04}, 5) \cite{McArtetal01}, 
6) \cite{Beuetal2003}, 7) \cite{Beuetal2004}, 8) \cite{Zachetal04}, 9) \cite{Kisletal99}, 
10) \cite{Klemolatal87}, 11) \cite{Adelman-McCarthyetal06}, 12) \cite{Girardetal04}, 13) \cite{LH65}, 
14) \cite{Kharchenko01}, 15) \cite{Hansonetal04}, 16) \cite{TT96}, 17) \cite{vPaetal96}, 
18) \cite{Sheetsetal07}, 19) \cite{Stilletal95}, 20) \cite{Arenasetal00}, 21) \cite{Petersetal06}, 
22) \cite{Watsonetal95}, 23) \cite{RB94}, 24) \cite{Welshetal95}, 25) \cite{Casaresetal96b}, 
26) \cite{Glennetal94}, 27) \cite{Schetal97}, 28) \cite{Watsonetal03}, 29) \cite{Hoardetal98}, 
30) \cite{TT97}, 31) \cite{Bruch2003}, 32) \cite{Neustroev02}, 33) \cite{Thoroughgoodetal2004}, 
34) \cite{Schlegeletal86}, 35) \cite{Schwarzetal05}, 36) \cite{Harlaftisetal96}, 37) \cite{Huberetal98}, 
38) \cite{Potteretal04}, 39) \cite{Rodetal98}, 40) \cite{vanderHeydenetal02}, 41) \cite{Rodetal07b}, 
42) \cite{Tappertetal97}, 43) \cite{Ringwaldetal1996a}, 44) \cite{Sheetsetal05}, 45) \cite{TF03}, 
46) \cite{deMartinoetal95}, 47) \cite{Arenasetal98}, 48) \cite{Nogamietal99}, 49) \cite{Thor97}, 
50) \cite{Patetal04}, 51) \cite{Thor00}, 52) \cite{Marsh99}, 53) \cite{Morelas-Rueda03}, 
54) \cite{TF02}, 55) \cite{Thor04}, 56) \cite{Noretal02}, 57) \cite{Noretal00}, 58) \cite{Kafkaetal03}, 
59) \cite{Harveyetal95}, 60) \cite{TT01}, 61) \cite{Fiedleretal97}, 62) \cite{Unda-Sanzanaetal06}, 
63) \cite{Nayloretal05}, 64) \cite{Echevarriaetal07}, 65) \cite{Vandeetal03}, 66) \cite{Watsonetal03}, 
67) \cite{Vaytetetal07}, 68) \cite{TT00}, 69) \cite{Casaresetal96a}, 70) \cite{Ringwaldetal05}, 
71) \cite{Mickaelianetal02}, 72) \cite{Belletal05}, 73) \cite{Petersetal05}, 74) \cite{Stilletal94}, 
75) \cite{RW98}, 76) \cite{Szkodyetal03}, 77) \cite{Szkodyetal00}, 78) \cite{Shahbazetal96}, 
79) \cite{Mukaietla87}, 80) \cite{Wagneretal98}, 81) \cite{Hoardetal97}, 82) \cite{Ringwaldetal1996b}, 
83) \cite{Hoardetal96}, 84) \cite{Thor98}, 85) \cite{Skillmanetal95}, 86) \cite{Hoardetal00}, 
87) \cite{Ringwaldetal1994}, 88) \cite{Thoretal97}, 89) \cite{Staudeetal2001}, 90) \cite{Ringwald1994}, 
91) \cite{Wolfetal1998}, 92) \cite{Smak02}, 93) \cite{Marsh88}, 94) \cite{Tovmassianetal00}, 
95) \cite{Thor02b}, 96) \cite{Bianchinietal2001}, 97) \cite{Howetal06}, 98) \cite{Thoroughgoodetal2005}, 
99) \cite{Stilletal98}, 100) \cite{August97}, 101) \cite{Matsumotoetal00}, 102) \cite{Masonetal01}, 
103) \cite{Liebertetal82}, 104) \cite{Schetal91}, 105) \cite{Schetal93}, 106) \cite{Pattersonetal03}, 
107) \cite{Thor02a}, 108) \cite{Grootetal01}, 109) \cite{Skidmoreetal00}, 110) \cite{Steeghsetal01}, 
111) \cite{Hartleyetal05}, 112) \cite{Rosenetal96}, 113) \cite{Schwarzetal98}, 114) \cite{Bonnet-Bidaudetal96}, 
115) \cite{Neustroevetal06}, 116) \cite{Dhillonetal94}, 117) \cite{Rolfeetal02a}, 118) \cite{Rolfeetal02b}, 
119) \cite{Wuetal01}, 120) \cite{Tappertetal01}, 121) \cite{MD96}, 122) \cite{Mennickentetal99}, 
123) \cite{Schetal00}, 124) \cite{Mukaietal86}, 125) \cite{Catalnetal99}, 126) \cite{Rodetal07}, 
127) \cite{Aungwerojwitetal02}, 128) \cite{Rodetal04}, 129) \cite{Tovmassianetal07}, 130) \cite{Rodetal05}, 
131) \cite{Withametal2007}, 132) \cite{Szkodyetal05}, 133) \cite{Szkodyetal04}, 134) \cite{Southworthetal06}, 
135) \cite{Zharikovetal06}, 136) \cite{Szkodyetal02}
}
\end{table*}

\subsection{Distances and proper motions}

First precise trigonometric parallaxes of the brightest CVs were obtained from 
{\it Hipparcos Satellite} \citep{Due99}. Trigonometric parallaxes of some CVs later 
were measured by using either Hubble Space Telescope's Fine Guidence Sensor 
\citep{McArtetal99,McArtetal01,Beuetal2003,Beuetal2004,Harretal04,Roetal07} or the ground-based 
observations \citep{Thor03,Thor08}, as well. However, number of systems with distances obtained 
from the trigonometric parallax method is only about 30. Nevertheless, there 
exist reliable distance prediction techniques for the systems whose trigonometric 
parallaxes do not exist. The distances were predicted using the period-luminosity-colours 
(PLCs) relation of \cite{Aketal07a}. Using {\it 2MASS} ($JHK_{s}$) photometric data, the PLCs 
relation of \cite{Aketal07a} for CVs is reliable and valid in the ranges 
$0.032$ $< P(d) \leq 0.454$, $-0.08 < (J-H)_{0} \leq 1.54$, 
$-0.03 < (H-K_{s})_{0} \leq 0.56$ and $2.0 < M_{J} < 11.7$ mag, 
which are well covering the present data of this study. For a detailed description of the 
method by the PLCs relation, see \cite{Aketal07a,Aketal08}. 

% Figure 1
\begin{figure}
\begin{center}
\includegraphics[scale=0.375, angle=0]{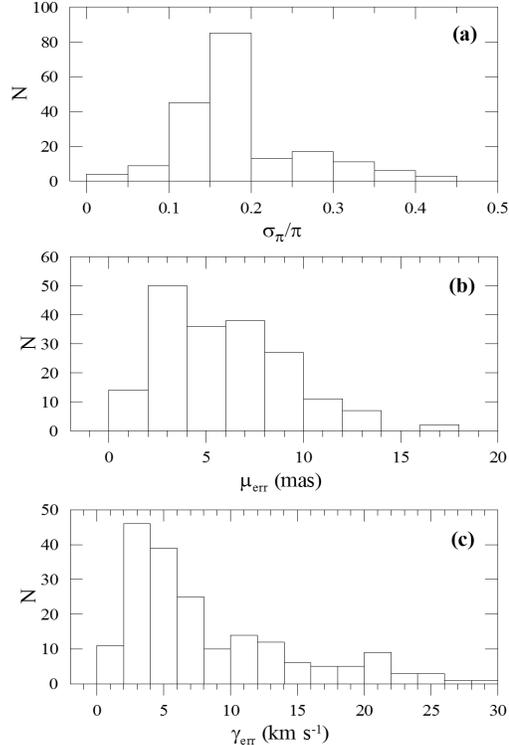}
\caption[] {\small Distribution of parallax errors (a), proper motion errors (b) and $\gamma$ 
velocity errors (c) of the present CV sample.}
\end{center}
\end{figure}

The proper motions of CVs were mostly taken from the {\it NOMAD} Catalogue of \cite{Zachetal04}. 
Distribution of parallax errors and proper motion errors of the present CV sample are shown 
in Fig. 1a-b, respectively. The median value and standard deviation of relative parallax errors 
are 0.16 and 0.08, respectively. The median value and standard deviation of proper motion errors 
are calculated 5.51 and 3.25 mas, respectively. Parallaxes and proper motion components are listed 
in Table 2 together with observational uncertainties. 

\subsection{Radial velocities}
Next necessary observational parameter to compute a space velocity of a star is its radial 
velocity with respect to Sun which comes from measurements of Doppler shifts of spectral lines. 
However, since CVs are binary stars their systemic velocities ($\gamma$) are used. vPAS96 have 
collected systemic velocities of CVs from literature covering times up to the year 1994. For 
this study, systemic velocities of CVs published in the literature up to the middle of 2007 were 
collected in a similar style. It has been a convention to use $V_{r}(\phi)=\gamma+K_{1,2}\sin\phi$ 
to predict systemic velocity of a CV from the measured radial velocities since orbits of CVs are 
circular, where $\phi$ is the orbital phase, $K_{1,2}$ are semi-amplitudes of radial velocity 
variation and 1 and 2 represent primary and secondary, respectively. $\gamma$ is the center 
of mass radial velocity of a CV, which is usually found by non-linear least-square fit of the 
function to observed radial velocities. 

The criteria defined by vPAS96 are adopted when collecting $\gamma$ velocities. Thus, if there are 
more than one determination of $\gamma$ for a CV, the mean of existing $\gamma$ values is taken. 
A new mean value is computed similarly if there is new $\gamma$ velocity measurement which is not 
in vPAS96. Sometimes, more than one $\gamma$ value obtained with different methods are encountered, 
then the value recommended by the author was adopted. The velocity measurements obtained during 
superoutbursts of SU UMa type dwarf novae have not been used since very large variations in 
$\gamma$ velocities have been observed during superoutbursts. 

In the sample, $\gamma$ velocities of 59 systems (32 dwarf novae, 25 nova like stars, 1 nova 
and 1 CV with unknown type) are from the absorption lines ($\gamma_{abs}$). Consequently, they 
are preferably used. However, it is well known that radial velocities derived from emission 
lines are likely affected by the motion in the accretion disc or the matter stream falling on 
the disc from the secondary. Thus, the measurements of $\gamma$ velocity obtained from emission 
lines ($\gamma_{em}$) may be unreliable \citep{Noretal02}. Radial velocities from the absorption 
lines of secondaries represent the system best. It should be noted that 
these very weak absorption lines are usually not observable in all CVs. They are most often visible 
in spectra of systems with orbital periods above the period gap \citep{Noretal02}. 

For this study, we too look for possible systematic errors in the $\gamma_{em}$ values and to provide 
information on the systematic and statistical accuracy of the $\gamma_{em}$ values. In our data 
sample there are 47 systems (28 dwarf novae, 18 nova like stars and 1 nova) with both the 
$\gamma_{em}$ and $\gamma_{abs}$ values, from which we find an average difference 
$<\gamma_{em}-\gamma_{abs}>=+0.57\pm14.3$ km s$^{-1}$, where the error is the standard deviation 
of the distribution of individual differences. This is a result which is in agreement with vPAS96 
who calculated an average difference $<\gamma_{em}-\gamma_{abs}>=+2.5\pm13.8$ km s$^{-1}$ using 
only 10 systems and concluded that there is no substantial systematic difference between systemic 
velocities derived from emission and absorption lines. Thus we too could conclude that the 
observed $\gamma$ velocities may allow a meaningful statistical analysis. Error histogram of 
$\gamma$ velocities is shown in Fig. 1c. The median value and standard deviation of $\gamma$ 
velocity errors are 5.6 and 6.3 km~s$^{-1}$, respectively. $\gamma$ velocities and associated 
uncertainties are listed in Table 2 together with distances and proper motion components. 

\subsection{Galactic space velocities}

Space velocities with respect to Sun were computed using the algorithms and 
transformation matrices of \cite{JS87}. The input data (celestial coordinates ($\alpha$, $\delta$), 
proper motion components ($\mu_{\alpha}\cos(\delta)$, $\mu_{\delta}$), systemic velocity $\gamma$ 
and the parallax $\pi$) are in the form adopted for the epoch of J2000 as described in the 
International Celestial Reference System (ICRS) of the Hipparcos and the Tycho Catalogues. 
Although sampled CVs are generally not distant objects, corrections for differential galactic 
rotation as described in \cite{MB81} were applied to the space velocities. 

The transformation matrices of \cite{JS87} use the notation of the right handed system. 
Therefore, the $U$, $V$, $W$ are the components of a velocity vector of a star with respect to 
the Sun, where the $U$ is directed toward the Galactic Center ($l=0^{o}, b=0^{o}$); the $V$ 
is in the direction of the galactic rotation ($l=90^{o}, b=0^{o}$); and the $W$ is towards 
the North Galactic Pole ($b=90^{o}$). 

% Figure 2
\begin{figure}
\begin{center}
\includegraphics[scale=0.5, angle=0]{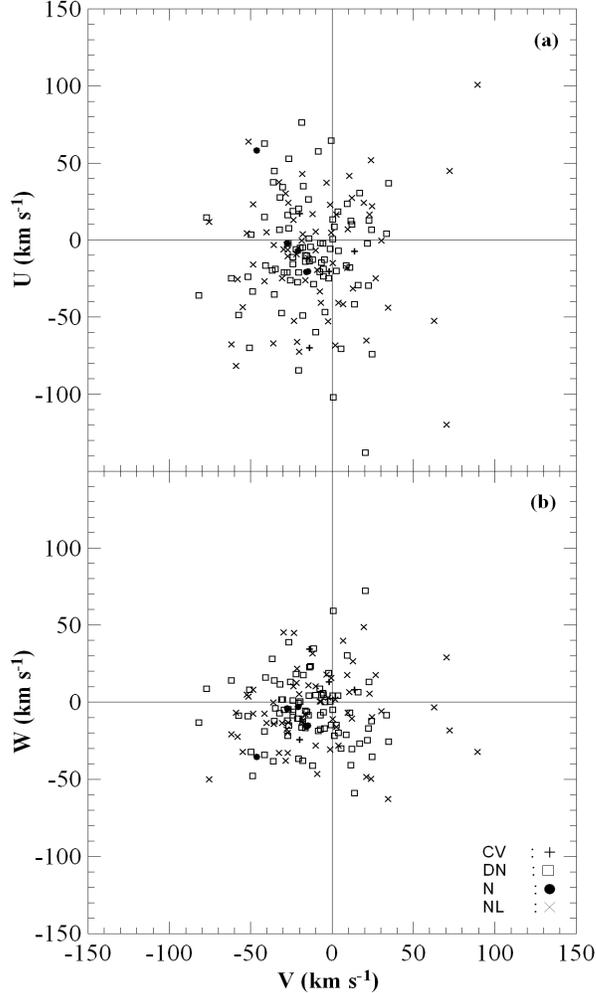}
\caption[] {\small Distribution of the space velocity components of CVs on the $U-V$ (a) and $W-V$ (b) 
planes. CV denotes CVs with unknown types, DN dwarf novae, N novae and NL nova-like stars.}
\end{center}
\end{figure}

The computed and corrected (galactic differential rotation) space velocity components are 
displayed in Fig. 2. Reliability of space velocities to indicate kinematical ages, the 
dispersion must be larger than the propagated errors. 
Therefore, the uncertainties of the space velocity components were also computed by propagating 
the uncertainties of the input data (proper motions, parallax and radial velocity) with 
an algorithm also by \cite{JS87}. The uncertainties of the space velocities 
($S_{err}^{2}=U_{err}^{2}+V_{err}^{2}+W_{err}^{2}$) were calculated, as well. Fig. 3 displays the 
histograms of the propagated uncertainties of space velocities ($S_{err}$), and their 
components ($U_{err}, V_{err}, W_{err}$). Apparently, very large uncertainties, which are 
bigger than $S_{err} > 30$ km s$^{-1}$, exist not only for space velocity vectors but also 
among the components. A trial test showed us that removing some CVs with the largest space 
velocity errors reduces the dispersions of the space velocity as well as dispersion of its components 
($U$, $V$, $W$). Thus, we have decided to refine the sample by removing all CVs with 
$S_{err} > 30$ km s$^{-1}$, which is the limit found by trial and error that shifting the $S_{err}$ 
limit further to a smaller value does not affect the space velocity dispersions. Thus, we have left 
159 systems in the refined sample.

The $S_{err}$ limit and refined sample are indicated in Fig. 3 where the unshaded areas are the 
error distribution of the whole sample, and the shaded areas are the error distribution of the 
refined sample. The median values of the errors for the refined sample are $\widetilde U_{err} = 7.7$, 
$\widetilde V_{err} = 7.5$ and $\widetilde W_{err} = 7.0$ km s$^{-1}$ while error distributions have 
standard deviations $\pm 4.8$, $\pm 4.4$ and $\pm 4.4$ km s$^{-1}$, respectively for the 
$U$, $V$ and $W$ components of the space velocity vectors. Such mean errors being smaller than 
dispersions on $U-V$ and $W-V$ diagrams (Fig. 2) indicate that the computed space velocity components 
have sufficient accuracy. 

% Figure 3
\begin{figure}
\begin{center}
\includegraphics[scale=0.375, angle=0]{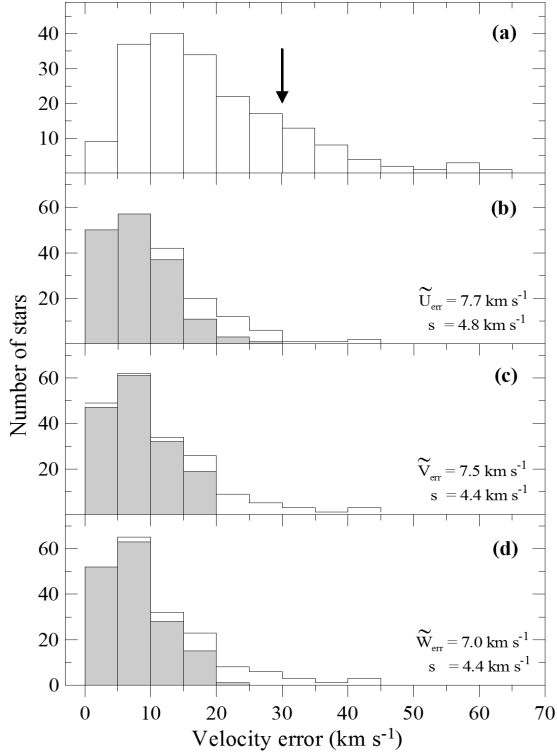}
\caption[] {\small Error histograms for space velocity vectors (a) and space velocity 
components (b-d) of CVs. The arrow indicates preferred $S_{err}$ limit.}
\end{center}
\end{figure}

Further test of reliability is demonstrated in Fig. 4a-b that some systems as 1) the most distant five systems, 
2) five systems with the biggest errors of space velocity vectors, and 3) five systems which are the 
most dispersed on the $U-V$ and $W-V$ planes. The position for the local standard of rest (LSR) is 
known by subtracting the Sun's velocity $(U, V, W)_{\odot}=(9,12,7)$ km s$^{-1}$ \citep{MB81} 
from the frame of ($U$, $V$, $W$) space velocities with respect to the Sun, and it is shown by a big 
outlined plus sign in Fig. 4a-b. One can see that the errors of the systems in the figure are not 
bigger than $\pm 26$ km s$^{-1}$ in both planes. Systems with biggest errors are not most dispersive 
ones. They are near the middle of the $U-V$ and $W-V$ planes. Fig. 4 indicates that dispersion in the 
velocity planes are unlikely to be affected by the velocity errors. Thus, the space velocities and the 
space velocity dispersions for the refined sample are reliable. 

% Figure 4
\begin{figure}
\begin{center}
\includegraphics[scale=0.5, angle=0]{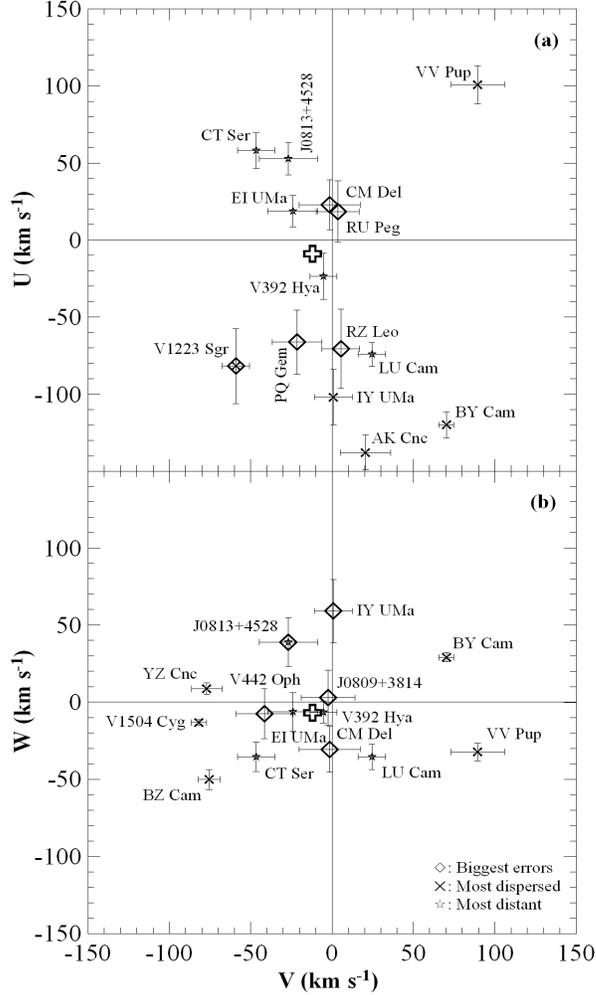}
\caption[] {\small The most distant five systems (stars), five systems with the biggest errors of 
space velocity vectors (diamonds), and five systems which are the most dispersed (times) on the 
$U-V$ and $W-V$ planes. The position of LSR is marked by a big-outlined plus symbol.}
\end{center}
\end{figure}

\subsection{Space distributions} 
Before interpreting space velocity dispersions, it is useful to investigate the space 
distribution of the sample in the solar neighbourhood. In order to inspect the spatial 
distribution of the refined sample, the Sun centered rectangular galactic coordinates 
($X$ towards Galactic Centre, $Y$ galactic rotation, $Z$ North Galactic Pole) were computed. 
The projected positions on the galactic plane ($X$, $Y$ plane) and on the plane perpendicular 
to it ($X$, $Z$ plane) are displayed in Fig. 5. Median distance of the whole sample is 
almost 300~pc which is well within the galactic disc in the solar neighbourhood. Fig. 5 shows 
that different sub-groups of CVs in the sample share almost the same space in the solar 
neighbourhood. It should be noted that \cite{Aketal08} studied a CV sample of 459 systems 
including the present refined sample and found a median distance of 377 pc. A comparison shows 
that the refined sample of this study occupies nearly the same space of the sample used 
by \cite{Aketal08} despite the number of stars is three times less. 

% Figure 5
\begin{figure}
\begin{center}
\includegraphics[scale=0.5, angle=0]{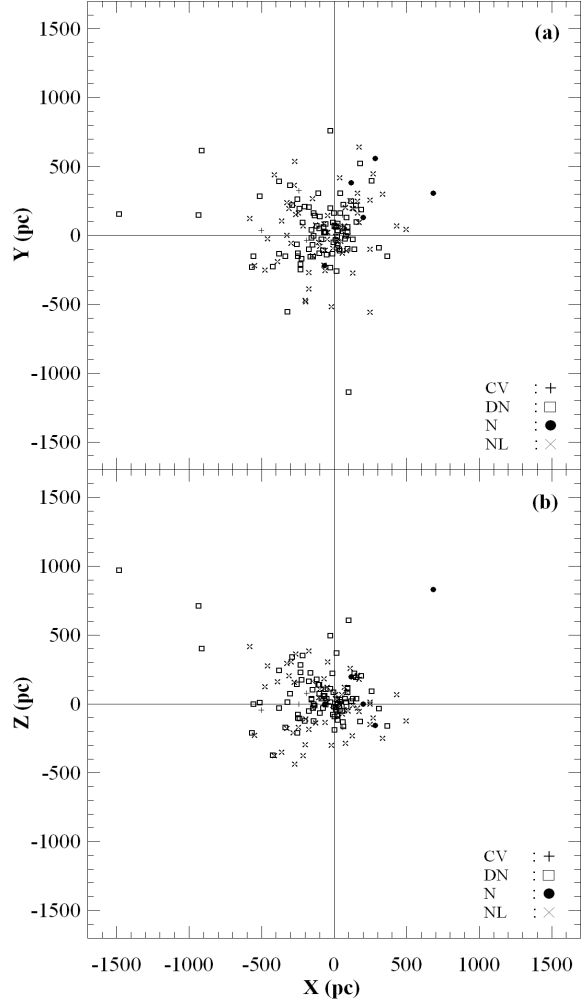}
\caption[] {\small The spatial distribution of CVs with respect to the Sun. 
$X$, $Y$ and $Z$ are heliocentric galactic coordinates directed towards the Galactic Centre, 
galactic rotation and the North Galactic Pole, respectively. Denotes are as in Fig. 3.}
\end{center}
\end{figure}

\subsection{Population analysis} 

Population analysis was also needed in this study when investigating kinematical ages of 
some sub-groups. There are essentially two ways to find population types (thin and thick discs 
or halo) of solar neighbourhood field stars; the pure kinematical approach, or by looking at 
a combination of kinematics, metallicities, and stellar ages \citep{Bensetal03}. Here, we adopt 
the pure kinematical approach. \cite{Bensetal03} suggested a method to minimize the contamination 
of thick-disc stars with thin-disc stars. Their method is based on the assumption that 
the galactic space velocities of the stellar populations with respect to the LSR in the thin 
disc, the thick disc and the halo have Gaussian distributions, 

\begin{equation}
f(U,~V,~W)=k~.~\exp\Biggl(-\frac{U_{LSR}^{2}}{2\sigma_{U{_{LSR}}}^{2}}-\frac{(V_{LSR}-V_{asym})
^{2}}{2\sigma_{V{_{LSR}}}^{2}}-\frac{W_{LSR}^{2}}{2\sigma_{W{_{LSR}}}^{2}}\Biggr),
\end{equation}
where 

\begin{equation}
k=\frac{1}{(2\pi)^{3/2}\sigma_{U{_{LSR}}}\sigma_{V{_{LSR}}}\sigma_{W{_{LSR}}}}
\end{equation}

normalizes the expression. $\sigma_{U{_{LSR}}}$, $\sigma_{V{_{LSR}}}$ and $\sigma_{W{_{LSR}}}$ 
are the characteristic velocity dispersions: 35, 20 and 16 km s$^{-1}$ for thin disc ($D$); 67, 38 
and 35 km s$^{-1}$ for thick disc ($TD$); 160, 90 and 90 km s$^{-1}$ for halo ($H$), respectively 
\citep{Bensetal03}. $V_{asym}$ is the asymmetric drift: -15, -46 and -220 km s$^{-1}$ for thin disc, 
thick disc and halo, respectively. $U_{LSR}$, $V_{LSR}$ and $W_{LSR}$ are galactic space velocities 
with respect to the LSR. Space velocity components with respect to LSR are obtained by adding space 
velocities of CVs to the space velocity of Sun ($(U, V, W)_{\odot}=(9,12,7)$ km s$^{-1}$, 
\cite{MB81}) with respect to LSR. 

In order to find the probability of a star belonging to a population, the probabilities from Eq. (1) 
has to be multiplied by the observed fractions ($X$) of each population in the solar neighbourhood. 
Observed fractions of thin-disc, thick-disc and halo stars in the solar neighbourhood are adopted 
$X_{D}=0.94$, $X_{TD}=0.06$ and $X_{H}=0.0015$, respectively \citep{Robetal96,Busetal99}. Then, two 
relative probabilities for the thick disc to thin disc ($TD/D$) and thick disc to halo ($TD/H$) 
membership for each star are estimated as following

\begin{equation}
TD/D=\frac{X_{TD}}{X_{D}}.\frac{f_{TD}}{f_{D}},~~~~~~~~~~TD/H=\frac{X_{TD}}{X_{H}}.\frac{f_{TD}}{f_{H}}.
\end{equation}

Stars are selected as from four different $TD/D$ intervals: $TD/D < 0.1$ (i.e. ``high probability 
thin-disc stars''); $0.1< TD/D < 1$ (i.e. ``low probability thin-disc stars''); $1< TD/D < 10$ 
(i.e. ``low probability thick-disc stars'') and $TD/D > 10$ (i.e. ``high probability thick-disc 
stars'') \citep{Bensetal05}.

$V_{LSR}-\log(TD/D)$ and Toomre diagrams of CVs in the refined sample are shown in Fig. 6. Horizontal 
lines divide the $V_{LSR}-\log(TD/D)$ diagram into four regions from bottom to the top; high 
probability thin-disc stars, low probability thin disc-stars, low probability thick-disc stars 
and high probability thick-disc stars. In the Toomre diagram of CVs, dashed circles delineate 
constant peculiar galactic space velocities $(U_{LSR}^{2}+W_{LSR}^{2})^{1/2}$ in steps of 
50 km s$^{-1}$. According to the criteria given by \cite{Bensetal05}, 131 systems ($\sim83\%$) 
in the present refined CV sample are high probability thin-disc stars, 18 systems ($\sim11\%$) low 
probability thin-disc stars, 5 systems ($\sim3\%$) low probability thick-disc stars and 5 systems 
($\sim3\%$) high probability thick-disc stars. Thus, it can be concluded that the CVs in this study 
mostly belong to the thin disc population. 

% Figure 6
\begin{figure}
\begin{center}
\includegraphics[scale=0.5, angle=0]{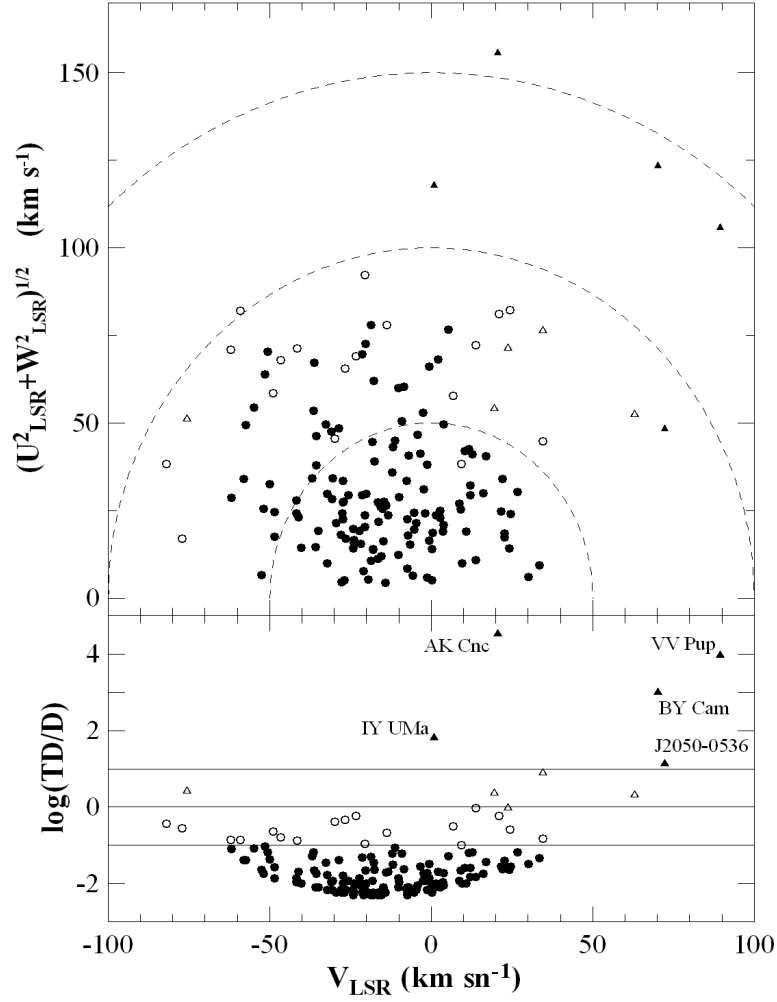}
\caption[] {\small {\it Lower panel}: $V_{LSR}-\log(TD/D)$ diagram of CVs. Filled and 
empty circles denote high and low probability thin-disc stars, while empty and filled 
triangles denote low and high probability thick-disc stars, respectively.
{\it Upper panel}: Toomre diagram of the same CV sample.}
\end{center}
\end{figure}

\subsection{Velocity dispersions and kinematical ages} 

Kinematical ages of CV samples were calculated from their velocity dispersions by the 
well known relation of \cite{Cox00}
\begin{equation}
\sigma_{\nu}^{3}(\tau)=\sigma_{\nu,\tau=0}^{3}+\frac{3}{2}\alpha_{V}\delta_{2}T_{\delta}
\Biggl[\exp\Biggl(\frac{\tau}{T_{\delta}}\Biggl)-1\Biggr],
\end{equation}
where, $\sigma_{\nu,\tau=0}$ is the velocity dispersion at zero age, which is usually taken 
as 10 km s$^{-1}$ \citep{Cox00}, $\alpha_{V}$ is a parameter describing the rotation 
curve ($\approx2.95$), $T_{\delta}$ is  a time scale ($5\times10^{9}$ yr), $\delta_{2}$ is 
a diffusion coefficient ($3.7\times10^{-6}$ (km~s$^{-1})^{3}$ yr). $\sigma_{\nu}(\tau)$ is 
the total velocity dispersion of the group of CVs (sample). $\tau$ is the kinematical age of 
the group, which is to be computed. The total dispersion of space velocity vectors 
($\sigma_{\nu}$) is connected to the dispersion of the velocity components as

\begin{equation}
\sigma_{\nu}^{2}=\sigma_{U{_{LSR}}}^{2}+\sigma_{V{_{LSR}}}^{2}+\sigma_{W{_{LSR}}}^{2}.
\end{equation}

After computing $\sigma_{\nu}^{2}$ from the dispersions of velocity components, it is used 
in Eq. (4) for computing $\tau$.

For an isotropic distribution the $\gamma$ velocity dispersion is defined \citep{Wieetal92,vPaetal96}

\begin{equation}
\sigma_{\gamma}^{2}=\frac{1}{3}\sigma_{\nu}^{2}.
\end{equation}

The $\gamma$ velocity dispersions are computed for a direct comparison to its theoretical predictions 
according to standard evolution model by \cite{KS96} and \cite{Kol01}, and listed in Table 3 for several 
sub-groups of CVs.

%Table 3

\begin{landscape}  
\begin{table*}
\setlength{\tabcolsep}{2.1pt} 
\begin{center}
\scriptsize{
\caption{Mean space velocities, space velocity dispersions, kinematical ages and $\gamma$ velocity 
dispersions ($\sigma_{\gamma}$) of CVs. $TD/D < 0.1$ denotes high probability thin disc, 
$0.1 < TD/D \leq 1$ low probability thin disc and $TD/D > 1$ low and high probability thick disc. 
Nova-like stars in the table do not include magnetic systems (polars and intermediate 
polars). N is the number of systems.}
\begin{tabular}{lcccccccccc}
\hline
Parameter & $N$ & $<U_{LSR}>$ & $<V_{LSR}>$ & $<W_{LSR}>$ & $\sigma_{U{_{LSR}}}$ & $\sigma_{V{_{LSR}}}$ & 
$\sigma_{W{_{LSR}}}$ & $\sigma_{\nu}$ & Age & $\sigma_{\gamma}$   \\
   &   &(km s$^{-1}$) & (km s$^{-1}$) & (km s$^{-1}$) & (km s$^{-1}$) & (km s$^{-1}$) & (km s$^{-1}$) & (km s$^{-1}$) & (Gyr)  & (km s$^{-1}$) \\
\hline
All sample & 159 & $+$0.07$\pm$2.93 & $-$0.48$\pm$2.19 & $+$1.53$\pm$1.80 & 36.86$\pm$4.78 & 27.48$\pm$4.45 & 22.69$\pm$3.65 & 51.27$\pm$7.48 & 4.84$\pm$1.37 & 30$\pm$4   \\
\hline
Dwarf novae & 87 & $+$0.41$\pm$3.81 & $-$2.36$\pm$2.52 & $+$1.59$\pm$2.35 & 35.31$\pm$4.82 & 23.46$\pm$4.42 & 21.84$\pm$4.43 & 47.69$\pm$7.90 & 4.19$\pm$1.44 & 28$\pm$5  \\
Nova-like stars & 38 & $+$1.67$\pm$5.20 & $-$1.63$\pm$4.72 & $+$0.97$\pm$4.07 & 31.67$\pm$3.86 & 28.74$\pm$4.65 & 24.79$\pm$4.36 & 49.43$\pm$7.45 & 4.51$\pm$1.36 & 29$\pm$4 \\
Novae & 5 & $+$10.49$\pm$14.59 & $-$13.16$\pm$5.81 & $-$7.64$\pm$5.87 & 31.00$\pm$3.23 & 17.56$\pm$3.38 & 14.01$\pm$2.13 & 38.28$\pm$5.14 & 2.57$\pm$0.86 & 22$\pm$3 \\
\hline
Magnetic systems & 25& $-$3.82$\pm$9.97 & $+$9.19$\pm$7.78 & $+$1.87$\pm$4.71 & 49.01$\pm$5.78 & 39.20$\pm$4.05 & 23.15$\pm$4.52 & 66.89$\pm$8.38 & 7.68$\pm$1.44 & 39$\pm$5 \\
\hline  
Non-magnetic systems & 134& $+$0.80$\pm$2.96 & $-$2.29$\pm$2.13 & $+$1.46$\pm$1.96 & 34.11$\pm$4.57 & 24.68$\pm$4.53 & 22.61$\pm$4.41 & 47.79$\pm$7.80 & 4.21$\pm$1.42 & 28$\pm$5 \\
\hline  
$TD/D < 0.1$ & 131 & $+$2.97$\pm$2.49 & $-$2.37$\pm$1.81 & $+$1.34$\pm$1.46 & 28.60$\pm$4.68 & 20.76$\pm$4.48 & 16.65$\pm$4.39 & 39.07$\pm$7.83 & 2.70$\pm$1.35 & 23$\pm$5 \\
$0.1 < TD/D \leq 1$ & 18 & $-$13.59$\pm$12.09 & $-$11.41$\pm$8.43 & $+$0.24$\pm$8.66 & 51.67$\pm$5.64 & 36.57$\pm$3.91 & 35.69$\pm$4.73 & 72.67$\pm$8.33 & 8.68$\pm$1.38 & 42$\pm$5 \\
$TD/D > 1$ & 10 & $-$13.34$\pm$25.60 & $+$43.88$\pm$15.00 & $+$6.24$\pm$15.73 & 77.95$\pm$4.30 & 62.85$\pm$4.87 & 47.59$\pm$4.71 & 110.87$\pm$8.02 & 14.35$\pm$0.99 & 64$\pm$5 \\
\hline  
$P(h) < 2.62$ & 62 & $+$0.46$\pm$4.82 & $+$1.38$\pm$3.89 & $+$8.29$\pm$3.16 & 37.64$\pm$5.10 & 30.45$\pm$4.41 & 26.02$\pm$4.53 & 54.96$\pm$8.12 & 5.52$\pm$1.48 & 32$\pm$5 \\
$P(h)\geq2.62$ & 97 & $-$0.18$\pm$3.71 & $-$1.67$\pm$2.59 & $-$2.80$\pm$2.05 & 36.35$\pm$4.50 & 25.40$\pm$4.42 & 20.28$\pm$4.22 & 48.76$\pm$7.59 & 4.39$\pm$1.39 & 28$\pm$4 \\
\hline  
$1.20<P(h)\leq1.83$ & 40 & $-$6.10$\pm$5.98 & $-$0.78$\pm$4.84 & $+$5.30$\pm$3.75 & 37.83$\pm$5.50 & 30.23$\pm$5.10 & 23.98$\pm$4.65 & 54.04$\pm$8.83 & 5.35$\pm$1.61 & 31$\pm$5 \\
$1.83<P(h)\leq3.35$ & 39 & $+$5.39$\pm$5.30 & $+$3.45$\pm$4.51 & $+$4.82$\pm$3.84 & 32.85$\pm$4.34 & 27.99$\pm$4.12 & 24.18$\pm$4.61 & 49.47$\pm$7.55 & 4.51$\pm$1.38 & 29$\pm$4 \\
$3.35<P(h)\leq4.50$ & 39 & $-$2.01$\pm$6.55 & $-$1.12$\pm$4.66 & $-$3.67$\pm$3.70 & 40.41$\pm$4.47 & 28.72$\pm$4.15 & 23.07$\pm$3.89 & 54.68$\pm$7.23 & 5.47$\pm$1.32 & 32$\pm$4 \\
$4.50<P(h)\leq11.00$ & 39 & $+$1.95$\pm$5.90 & $-$4.39$\pm$3.56 & $-$1.83$\pm$2.99 & 36.44$\pm$4.69 & 22.40$\pm$4.51 & 18.54$\pm$4.23 & 46.62$\pm$7.76 & 4.00$\pm$1.41 & 27$\pm$4 \\
\hline
\end{tabular}
}
\end{center}
\end{table*}
\end{landscape}  

%Table 4
\begin{landscape}  
\begin{table*}
\setlength{\tabcolsep}{2.1pt} 
\begin{center}
\scriptsize{
\caption{Mean space velocities, space velocity dispersions, kinematical ages and $\gamma$ velocity 
dispersions ($\sigma_{\gamma}$) of non-magnetic systems according to the orbital period 
ranges. N is the number of CVs.}
\begin{tabular}{lcccccccccc}
\hline
Parameter & $N$ & $<U_{LSR}>$ & $<V_{LSR}>$ & $<W_{LSR}>$ & $\sigma_{U{_{LSR}}}$ & $\sigma_{V{_{LSR}}}$ & 
$\sigma_{W{_{LSR}}}$ & $\sigma_{\nu}$ & Age & $\sigma_{\gamma}$   \\
   &   &(km s$^{-1}$) & (km s$^{-1}$) & (km s$^{-1}$) & (km s$^{-1}$) & (km s$^{-1}$) & (km s$^{-1}$) & (km s$^{-1}$) & (Gyr)  & (km s$^{-1}$) \\
\hline
$P(h) < 2.62$ & 53 & $-$2.47$\pm$4.98 & $-$2.25$\pm$3.73 & $+$8.10$\pm$3.50 & 35.97$\pm$5.22 & 26.99$\pm$4.33 & 26.49$\pm$4.45 & 52.19$\pm$8.11 & 5.01$\pm$1.48 & 30$\pm$5 \\
$P(h)\geq2.62$ & 81 & $+$2.94$\pm$3.66 & $-$2.31$\pm$2.56 & $-$2.89$\pm$2.17 & 32.84$\pm$4.06 & 23.04$\pm$4.59 & 19.65$\pm$4.22 & 44.67$\pm$7.44 & 3.65$\pm$1.34 & 26$\pm$4 \\
\hline 
$1.06<P(h)\leq1.80$ & 34 & $-$6.60$\pm$5.81 & $-$3.23$\pm$4.61 & $+$6.88$\pm$4.22 & 34.04$\pm$4.36 & 26.68$\pm$4.72 & 25.21$\pm$4.79 & 50.06$\pm$8.01 & 4.62$\pm$1.47 & 29$\pm$5 \\
$1.80<P(h)\leq3.35$ & 33 & $+$2.19$\pm$6.06 & $+$0.38$\pm$4.70 & $+$3.41$\pm$4.20 & 34.35$\pm$5.23 & 26.57$\pm$4.46 & 24.01$\pm$4.43 & 49.62$\pm$8.18 & 4.54$\pm$1.50 & 29$\pm$5 \\
$3.35<P(h)\leq4.60$ & 33 & $+$0.90$\pm$6.23 & $-$3.65$\pm$4.33 & $-$2.33$\pm$3.66 & 35.23$\pm$3.78 & 24.76$\pm$4.38 & 20.84$\pm$4.01 & 47.84$\pm$7.04 & 4.22$\pm$1.29 & 28$\pm$4 \\
$4.60<P(h)\leq11.00$ & 33 & $+$6.17$\pm$5.73 & $-$2.68$\pm$3.60 & $-$3.41$\pm$3.32 & 33.01$\pm$4.42 & 20.52$\pm$4.61 & 19.09$\pm$4.14 & 43.30$\pm$7.61 & 3.41$\pm$1.36 & 25$\pm$4 \\
\hline     
\end{tabular}
}
\end{center}
\end{table*}
\end{landscape}  

\section{Discussions}

Except four systems (V392 Hya, CT Ser, EI UMa and J0813+4528) all of the refined sample 
are contained within 500 pc from the galactic plane. Thus, all of the CVs in the refined 
sample could be considered within the galactic disc (Fig. 5). Concentration towards the 
Sun and sparseness outwards indicate that space distribution of present data is highly 
affected by selection with respect to brightness. 

Various sub-groups of the refined sample have been formed to investigate if there are 
kinematically homogeneous sub-systems. All sample and outstanding sub-groups have been 
summarized together with their mean space velocity components, velocity dispersions of 
the components, total dispersion and corresponding kinematical age of the group according 
to Eq. (4) in Tables 3 and 4. The velocity dispersions in Table 3
($\sigma_{U{_{LSR}}}$ = 36.86 km s$^{-1}$, $\sigma_{V{_{LSR}}}$ = 27.48 km s$^{-1}$, 
$\sigma_{W{_{LSR}}}$ = 22.69 km s$^{-1}$) indicate that the present refined sample has 
a mean kinematical age of 4.84$\pm$1.37 Gyr. Considering the total dispersion 
($\sigma_{\nu}$=51.27 km s$^{-1}$) for the refined sample as a whole, its empirical 
$\gamma$ velocity dispersion is $\sigma_\gamma=30\pm4$ km s$^{-1}$ according to Eq. (6).

\subsection{Kinematics of CV types}
Kinematical properties of standard sub-groups; dwarf novae (DN), nova-like systems (NL) 
and novae were investigated first. Those sub-groups were shown with different symbols on 
the $U-V$ and $W-V$ diagrams in Fig. 2. Their kinematics were summarized in Table 3. 
There is no clear distinction between the distributions of first two sub-groups on the 
velocity space. Mean space velocities and velocity dispersions of dwarf novae 
and nova-like stars appear similar with slightly older kinematical age of nova-like 
stars. Although their number is statistically insignificant, novae appear to be the 
youngest ($2.57\pm0.86$ Gyr). Note that \cite{Aketal08} found novae in their sample as located 
almost on the galactic plane, suggesting that they are young systems. 

Comparisons of $U-V$ diagrams of dwarf novae, nova-like stars and novae are shown in 
Fig. 7. It should be noted that magnetic systems (polars and intermediate polars) 
are removed from the sample since the evolution of magnetic systems can be different than 
the evolution of non-magnetic CVs \citep{Wu1993,WebbinkWickramasinghe2002,Schwarzetal07}. 
Central concentration of considerable amount of dwarf novae towards LSR makes them 
appear slightly younger while a central concentration in the $U-V$ diagram 
of nova-like stars is not observed.

% Figure 7
\begin{figure}
\begin{center}
\includegraphics[scale=0.68, angle=0]{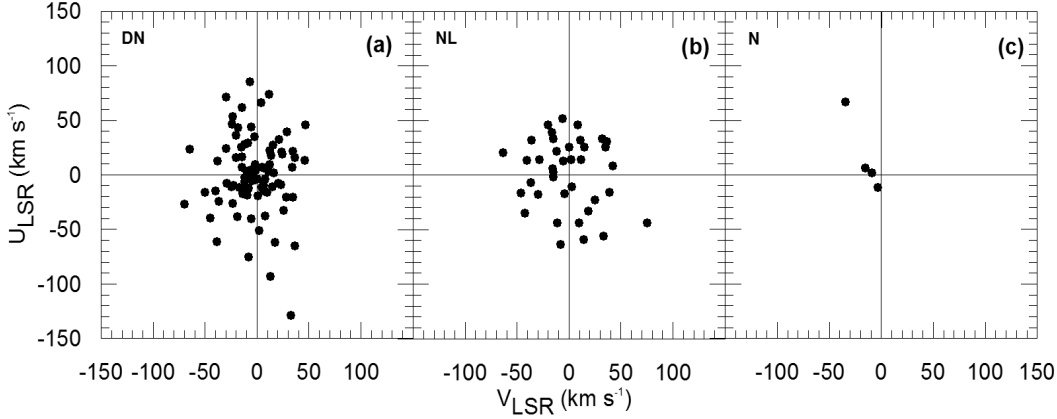}
\caption[] {\small $U-V$ diagrams of CVs for sub-types. DN denotes dwarf novae (a), 
NL nova-like stars (b) and N novae (c). Magnetic systems are not included in the sample.}
\end{center}
\end{figure}

Magnetic and non-magnetic systems display different kinematical properties (Table 3). 
Although the number of magnetic systems is small (N = 25) which is near the limit of 
statistical significance, they appear almost twice older than non-magnetic systems. 
Concentration towards LSR which makes the dispersion smaller for non-magnetic systems 
is clear in Fig. 8b. However, $\gamma$ velocities derived for magnetic CVs may be 
significantly contaminated by the flow velocities of the magnetically channelled plasma. 
Thus, their $\gamma$ velocities may not be reliable. Consequently, kinematically older appearance 
may not be the truth. 

% Figure 8
\begin{figure}
\begin{center}
\includegraphics[scale=0.48, angle=0]{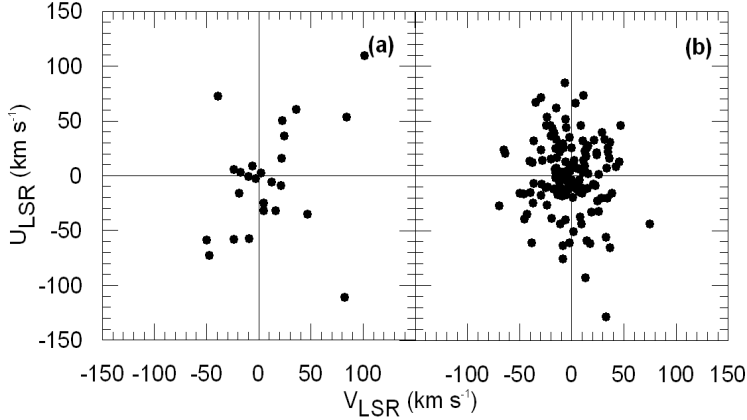}
\caption[] {\small $U-V$ diagrams of magnetic (a) and non-magnetic CVs (b).}
\end{center}
\end{figure}

\subsection{Groups according to population analysis}

Since population analysis is directly related to kinematical properties of stars, the 
distribution of the groups according to the population analysis on the $U-V$ diagram show 
very distinct properties (see Fig. 9 and Table 3). It is inconsistent to assign a kinematical 
age to these groups, because formation of groups is already according to their 
kinematics. 

% Figure 9
\begin{figure}
\begin{center}
\includegraphics[scale=0.68, angle=0]{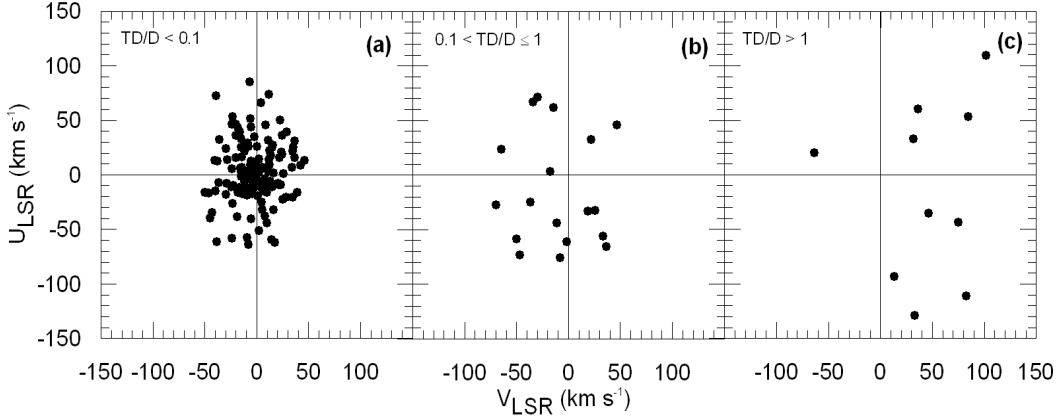}
\caption[] {\small $U-V$ diagrams of CV populations. $TD/D < 0.1$ shows high probability 
thin-disc stars (a), while $0.1 < TD/D \leq 1$ shows low probability thin-disc stars 
(b), and $TD/D>1$ represents both low and high probability thick-disc stars (c).}
\end{center}
\end{figure}

Among the 131 high probability thin-disc ($TD/D < 0.1$) CVs, the most are dwarf novae, 
with 33 SU UMa type stars, 21 U Gem type stars, 12 Z Cam type stars and nine unknown 
type dwarf novae. All Z Cam type dwarf novae are found among the high probability 
thin-disc stars. Among high probability thin-disc CVs, four are novae and three are 
unknown type. Number of nova-like stars in high probability thin-disc CVs is 49, with one AC Cnc type 
star, eight polars, nine intermediate polars, 16 UX UMa type stars, five VY Scl type stars 
and one unknown type nova-like star. Although magnetic systems appear to be old in Fig. 8, 
17 of them are contained in high probability thin disc group in Fig. 9 which indicates youth. 

Among the low probability thin-disc ($0.1 < TD/D \leq 1$) CVs, the 10 of them are dwarf 
novae (four SU UMa type stars, four U Gem type stars and two unknown type dwarf novae), six 
of them are nova-like stars (one polar, two intermediate polars, two UX UMa type stars and 
one unknown type nova-like star), one is nova and one is unknown type. 

Nova-like stars comprise the majority of low and high probability thick-disc stars ($TD/D > 1$). 
Among 10, eight of them are nova-like stars, with four polars, one intermediate polar, one 
VY Scl star and two unknown type. Only two of 10 are dwarf novae. 

In this study we find that almost all of CVs in the refined sample could be considered 
within the galactic disc (Fig. 5). Although there are relatively a few thick-disc and 
halo stars in solar neighbourhood, thick-disc and halo stars are dominant in $z$-distance 
intervals $1.5<z\leq5$, and $z>5$ kpc, respectively \citep{Kara04,Bil06,Aketal07b,Bil08}. 
Here, $z$ denotes vertical distance from the galactic plane. Thus, thick-disc and halo CVs 
must be the most distant systems. They can be detected by using the deep photometric 
systems, e.g. SDSS, IPHAS, etc. 

\subsection{Groups according to orbital periods}

Since the orbital period of a CV is directly related to the mean mass density of the secondary 
star, which is supposed to be an age related quantity, investigating kinematical properties 
according to orbital period ranges could be the most meaningful. Moreover, orbital period 
evolution of CV samples has been predicted by various authors (see introduction).

In order to search kinematical differences among the sub-groups of limited orbital period 
ranges, first, distribution of orbital periods must be studied. The orbital period distribution 
of the refined sample is shown in Fig. 10 where the orbital period gap is distinctly visible 
between 2 h and 3 h. Separating point of the period gap is found at 2.62 h by fitting Gaussians 
with two peaks to the orbital period distribution in Fig. 10. It should be emphasized that 
this approach is just one way to define the middle of the gap. 

% Figure 10
\begin{figure}
\begin{center}
\includegraphics[scale=0.46, angle=0]{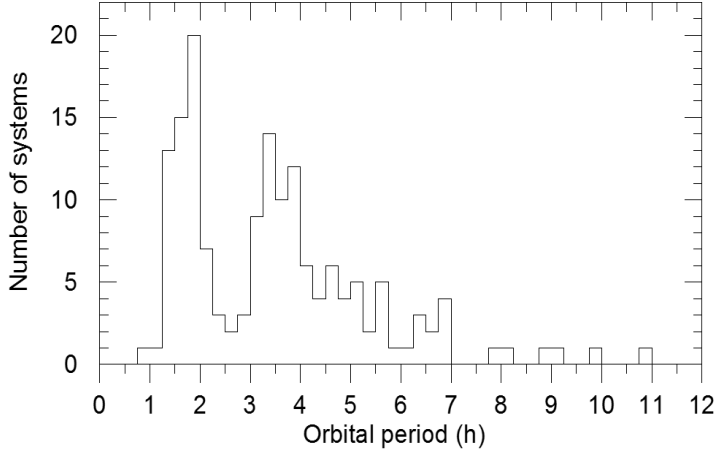}
\caption[] {\small Orbital period distribution of CVs in the refined sample.}
\end{center}
\end{figure}

The $U-V$ diagrams of the sub-groups of $P(h)<2.62$ and $P(h)\geq2.62$ are shown in Fig. 11 
and their kinematics are given in Table 3. Table 3 indicate that systems below the period gap 
($P(h)<2.62$) are kinematically older than systems above the gap ($P(h)\geq2.62$) as predicted 
by standard model. However, a comparison between the two distributions indicates $U-V$ diagram 
and other kinematical quantities are almost the same for these sub-groups. Similarity is indicated 
by that both groups have central concentrations (low dispersion stars) as well as high velocity 
stars. If one considers kinematical ages are the same within the error limits, it can be concluded 
that the result would be in agreement with vPAS96.

% Figure 11
\begin{figure}
\begin{center}
\includegraphics[scale=0.48, angle=0]{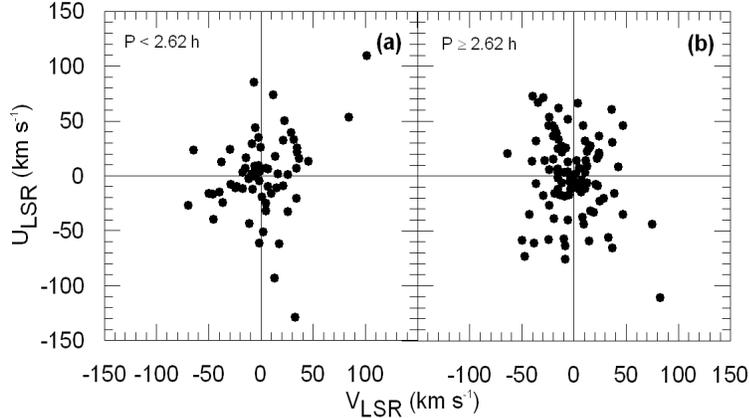}
\caption[] {\small $U-V$ diagrams of CVs below (a) and above (b) the orbital period 
gap.}
\end{center}
\end{figure}

Another way to investigate different period groups is to divide the sample according 
to the same number of CVs at different period ranges. Such a grouping is made and summarized 
in the last four lines of Table 3. Table 3 shows that there is an excess of the space velocity 
dispersion in $3.35<P(h)\leq4.50$ period interval. This is mainly due to magnetic systems 
(polars and intermediate polars) whose $\gamma$ velocities may not be reliable, as pointed 
above. Thus, magnetic systems are removed from the refined sample, leaving 134 non-magnetic 
CVs in the sample.

After removing the magnetic systems, the sample analyzed in a similar style. When comparing 
the kinematics of non-magnetic CVs below and above the period gap (Table 4 and Fig. 12), it has 
been found that difference of kinematical ages slightly incereased as indicated by dispersions. 
On the other hand, kinematics and $U-V$ diagrams of the four sub-groups without magnetic systems 
are summarized in the last four lines of Table 4. It becomes clear after removing magnetic systems 
from the sample that towards the longer periods dispersions become smaller with younger ages. 
The $U-V$ diagrams of four sub-groups of non-magnetic systems are shown in Fig. 13. It should 
be noted that for the systems grouped in Fig. 13 $\gamma$ velocity is increasingly difficult 
to measure from panel (d) to (a).

Table 4 shows that the mean kinematical age difference between the non-magnetic systems above 
and below the orbital period gap is about 1.4 Gyr. Thus, kinematics of the present sample of CVs 
roughly confirm the prediction of \cite{KS96} who predicted about 2 Gyr age difference between the CV groups 
of below and above the period gap. 

% Figure 12
\begin{figure}
\begin{center}
\includegraphics[scale=0.48, angle=0]{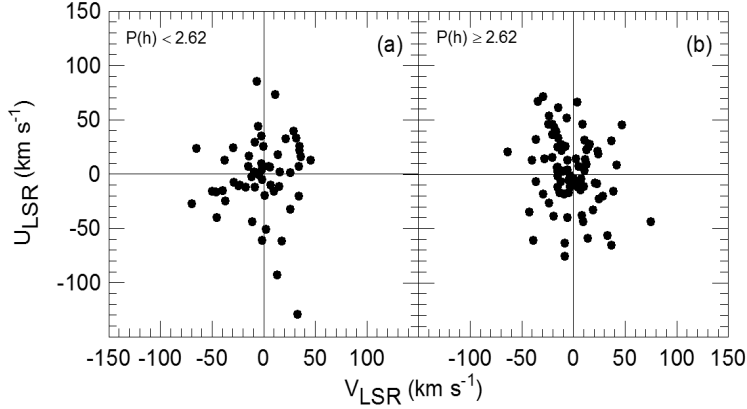}
\caption[] {\small $U-V$ diagrams of non-magnetic CVs below ($P(h)<$2.62) 
and above ($P(h)\geq2.62$) the orbital period gap.}
\end{center}
\end{figure}

% Figure 13

\begin{figure}
\begin{center}
\includegraphics[scale=0.48, angle=0]{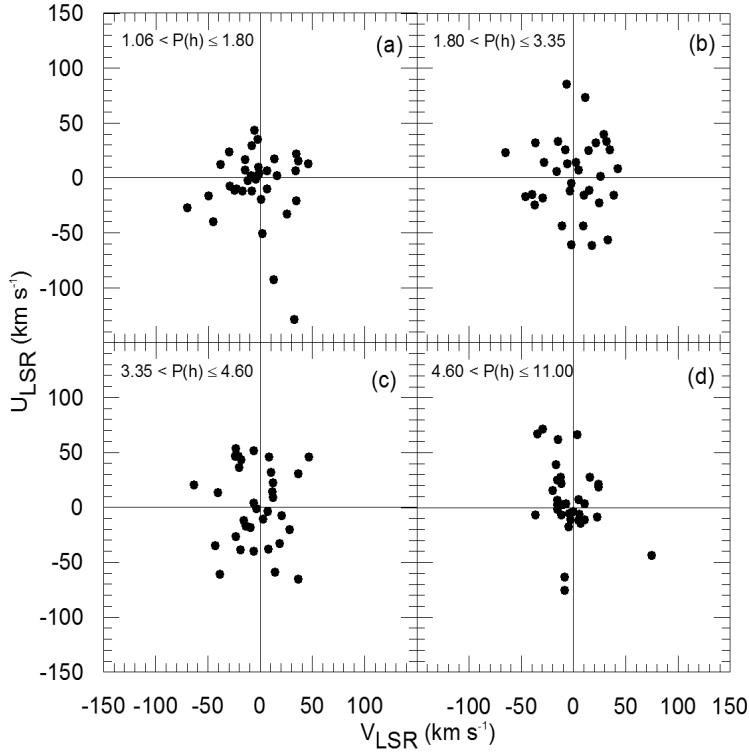}
\caption[] {\small $U-V$ diagrams of non-magnetic CVs in different orbital period ranges.}
\end{center}
\end{figure}

\subsection{$\gamma$ velocity dispersion of CVs}

The definition of $\gamma$ velocity dispersion in Eq. (6) is based on the assumption 
that the distribution of space velocity components is isotropic \citep{Wieetal92,vPaetal96}. 
However, the truth is that it is not isotropic as $\sigma_{U{_{LSR}}}$ is always higher 
than $\sigma_{V{_{LSR}}}$ and $\sigma_{W{_{LSR}}}$ (see Table 3 and 4). Nevertheless, 
$\gamma$ velocity dispersion of a CV sub-group calculated from total space velocity dispersion 
($\sigma_{\nu}$) as in Eq. (6) is useful for direct comparison of theoretical predictions. 

The discussions of $\gamma$ velocity dispersions are based on non-magnetic systems. 
$\gamma$ velocity dispersion of all non-magnetic systems in the refined sample is estimated 
$\sigma_\gamma=28\pm5$ km s$^{-1}$. For a comparison with theoretical predictions, $\gamma$ 
velocity dispersions of non-magnetic CVs below ($P(h)<2.62$) and above ($P(h)\geq2.62$) the 
period gap are calculated $\sigma_\gamma=30\pm5$ km s$^{-1}$ and $\sigma_\gamma=26\pm4$ km s$^{-1}$ 
(see Table 4), respectively. Theoretical predictions of $\gamma$ velocity dispersion are 
$\sigma_\gamma\simeq30$ km s$^{-1}$ and $\sigma_\gamma\simeq15$ km s$^{-1}$ \citep{KS96} 
for below and above the gap, respectively. Although the observational $\gamma$ velocity 
dispersion of the systems below the period gap is in agreement with its theoretical prediction 
given by \cite{KS96}, the observational $\gamma$ velocity dispersion of the systems above 
the gap is not in agreement with its theoretical prediction. However, \cite{Kol01} suggested 
that the $\gamma$ velocity dispersions of CVs are $\simeq 32$ km~s$^{-1}$ below the period 
gap versus $\simeq 27$ km~s$^{-1}$ above the gap, respectively, if magnetic braking does not 
operate in the detached phase, during which the system evolves from the post-CE orbit into 
contact. It seems that our results confirm \cite{Kol01}.

It should be noted that cumulative effect of repeated nova explosions 
with possibly asymmetric envelope ejection would increase the $\gamma$ velocity dispersions of 
CVs during their evolution \citep{KS96}. In this case, the observational $\gamma$ velocity 
dispersions in this study would represent upper limits while theoretical values would 
be underestimated. Thus, difference between observational and theoretical values increases. 
In addition, if nova explosions increased the $\gamma$ velocity dispersions of CVs during their 
evolution, kinematical ages found in our study would also represent upper limits. However, nova 
explosions can not affect the age and $\gamma$ velocity dispersion differences between 
sub-groups of CVs. 

\section{Conclusions}

The $\gamma$ velocities of CVs are taken from a variety of sources from the literature. 
For all but 59 systems, published $\gamma$ velocities are by-products of measurements of the 
components' semi-amplitude. In addition, measurements based on emission lines remain problematic. 
As a consequence, systematic errors affect specifically the determination of $\gamma$ velocity 
and error values are high. 

After analysing available kinematical data of CVs, it is concluded that there is not 
considerable kinematical difference between dwarf novae and non-magnetic nova-like stars. 
Magnetic and non-magnetic systems display different kinematical properties. However, kinematics 
based on $\gamma$ velocity measurements of magnetic systems remain problematic as their 
$\gamma$ velocities may be significantly contaminated by the flow velocities of the 
magnetically channelled plasma.
 
Kinematics of the present sample show that systems above the orbital pediod gap are 
younger than systems below the gap. This result is in agreement with the standard theory 
of the CV evolution. Kinematics of the present sample of CVs also roughly confirm the 
prediction of \cite{KS96} who predicted about 2 Gyr age difference between the CV groups 
of below and above the period gap. Smaller age difference found in this study shows a similarity 
of the angular momentum loss time scales in systems with low-mass and high-mass secondary 
stars \citep{Kol01}. 

Assuming an isotropic distribution, observational $\gamma$ velocity dispersions of 
CVs are in agreement with the theoretical predictions of \cite{Kol01} who suggested that 
the difference of $\gamma$ velocity dispersions of the systems below and above the period 
gap is about 5 km~s$^{-1}$ with systems above the gap having a $\gamma$ velocity 
dispersion of $\simeq 27$ km~s$^{-1}$. This agreement between the observations and the 
theory implies that magnetic braking does not operate in the detached phase. 

\section{Acknowledgments}
Part of this work was supported by the Research Fund of the University of Istanbul, 
Project Number: BYP-1379. This research has made use of the SIMBAD database, operated at 
CDS, Strasbourg, France. This publication makes use of data products from the Two Micron 
All Sky Survey, which is a joint project of the University of Massachusetts and the Infrared 
Processing and Analysis Center/California Institute of Technology, funded by the National 
Aeronautics and Space Administration and the National Science Foundation. This research has 
made use of the NASA/IPAC Extragalactic Database (NED) which is operated by the Jet 
Propulsion Laboratory, California Institute of Technology, under contract with the National 
Aeronautics and Space Administration. We thank the anonymous referee for a thorough 
report and useful comments that helped improving an early version of the paper.

\end{document}